
\message{ File currently being modified--stay tuned for updated version.}
\message{*******************************************************************}
\newif\ifchapterpage
\newif\ifprelimpage
\newif\iftitlepage
\newif\ifnopagenumber
\newif\ifnopagenumbers

\newcount\currfam
\newcount\sqrfam 

\newskip\twelvebaseskip
\newskip\tenbaseskip
\newskip\footbaseskip
\newskip\vfootline
\newskip\vheadline

\font\TXrm=cmr10 scaled 1200 \font\qTXrm=cmr9 scaled 1095
   \font\qqTXrm=cmr9 scaled 900  \font\qqqTXrm=cmr7 scaled 900
\font\TXi=cmmi10 scaled 1200 \font\qTXi=cmmi9 scaled 1095
   \font\qqTXi=cmmi9 scaled 900  \font\qqqTXi=cmmi7 scaled 900
\font\TXsy=cmsy10 scaled 1200 \font\qTXsy=cmsy9 scaled 1095
   \font\qqTXsy=cmsy9 scaled 900 \font\qqqTXsy=cmsy7 scaled 900
\font\TXbf=cmbx10 scaled 1200 \font\qTXbf=cmbx9 scaled 1095
   \font\qqTXbf=cmbx9 scaled 900 \font\qqqTXbf=cmbx7 scaled 900
\font\TXit=cmti10 scaled 1200 \font\qTXit=cmti9 scaled 1095
   \font\qqTXit=cmti9 scaled 900 \font\qqqTXit=cmti7 scaled 900
\font\TXsl=cmsl10 scaled 1200 \font\qTXsl=cmsl9 scaled 1095
   \font\qqTXsl=cmsl9 scaled 900 \font\qqqTXsl=cmsl9 scaled 900
\font\TXtt=cmtt10 scaled 1200 \font\qTXtt=cmsl9 scaled 1095
\font\TXex=cmex10 scaled 1200  \font\qTXex=cmex10
\font\TXsc=cmcsc10 scaled 1200 \font\qTXsc=cmcsc10  

\skewchar\TXi='177 \skewchar\qTXi='177 \skewchar\qqTXi='177
  \skewchar\qqqTXi='177
\skewchar\TXsy='60 \skewchar\qTXsy='60 \skewchar\qqTXsy='60
  \skewchar\qqqTXsy='60


\def\twelvepoint{\def\rm{\currfam0 \sqrfam0 \fam0\TXrm}
  \textfont0=\TXrm \scriptfont0=\qTXrm \scriptscriptfont0=\qqTXrm
  \textfont1=\TXi  \scriptfont1=\qTXi  \scriptscriptfont1=\qqTXi
  \textfont2=\TXsy \scriptfont2=\qTXsy \scriptscriptfont2=\qqTXsy
  \textfont3=\TXex \scriptfont3=\TXex  \scriptscriptfont3=\TXex
  \textfont\itfam=\TXit \scriptfont\itfam=\qTXit
\scriptscriptfont\itfam=\qqTXit
    \def\it{\currfam\itfam \sqrfam0 \fam\itfam\TXit}%
  \textfont\slfam=\TXsl \scriptfont\slfam=\qTXsl
\scriptscriptfont\slfam=\qqTXsl
    \def\sl{\currfam\slfam \sqrfam0 \fam\slfam\TXsl}%
  \textfont\bffam=\TXbf \scriptfont\bffam=\qTXbf
\scriptscriptfont\bffam=\qqTXbf
    \def\bf{\currfam\bffam \sqrfam\bffam \fam\bffam\TXbf}%
  \textfont\ttfam=\TXtt \def\tt{\currfam\ttfam \sqrfam\ttfam \fam\ttfam\TXtt}%
  \normalbaselineskip=\twelvebaseskip \let\sc=\TXsc \normalbaselines%
  \setbox\strutbox=\hbox{\vrule height.7\baselineskip depth.3\baselineskip%
    width0pt}\rm}


\def\tenpoint{\def\rm{\currfam0 \sqrfam0 \fam0\qTXrm}
  \textfont0=\qTXrm \scriptfont0=\qqTXrm \scriptscriptfont0=\qqqTXrm
  \textfont1=\qTXi  \scriptfont1=\qqTXi  \scriptscriptfont1=\qqqTXi
  \textfont2=\qTXsy \scriptfont2=\qqTXsy \scriptscriptfont2=\qqqTXsy
  \textfont3=\qTXex \scriptfont3=\qTXex  \scriptscriptfont3=\qTXex
  \textfont\itfam=\qTXit \scriptfont\itfam=\qqTXit \scriptscriptfont\itfam=%
   \qqqTXit
    \def\it{\currfam\itfam \sqrfam0 \fam\itfam\qTXit}%
  \textfont\slfam=\qTXsl \scriptfont\slfam=\qqTXsl \scriptscriptfont\slfam=%
   \qqqTXsl
    \def\sl{\currfam\slfam \sqrfam0 \fam\slfam\qTXsl}%
  \textfont\bffam=\qTXbf \scriptfont\bffam=\qqTXbf \scriptscriptfont\bffam=%
   \qqqTXbf
    \def\bf{\currfam\bffam \sqrfam\bffam \fam\bffam\qTXbf}%
  \textfont\ttfam=\qTXtt \def\tt{\currfam\ttfam \sqrfam\ttfam
\fam\ttfam\qTXtt}%
  \normalbaselineskip=\tenbaseskip \let\sc=\qTXsc \normalbaselines%
  \setbox\strutbox=\hbox{\vrule height.7\baselineskip depth.3\baselineskip%
    width0pt}\rm}


\newfam\titlefam
  \font\titlef=cmbx10 scaled 1440
  \font\qtitlef=cmbx10 scaled 1200
  \font\qqtitlef=cmbx10 scaled 1000
  \textfont\titlefam=\titlef \scriptfont\titlefam=\qtitlef
  \scriptscriptfont\titlefam=\qqtitlef
  \def\titlefont{\currfam\titlefam \sqrfam\titlefam \fam\titlefam\titlef}

\newfam\BIGfam
  \font\BIGf=cmr9 scaled 1440
  \font\qBIGf=cmr9 scaled 1200
  \font\qqBIGf=cmr9 scaled 1000
  \textfont\BIGfam=\BIGf \scriptfont\BIGfam=\qBIGf
  \scriptscriptfont\BIGfam=\qqBIGf

\newfam\secfam
  \font\secf=cmbx9 scaled 1440
  \font\qsecf=cmbx9 scaled 1200
  \font\qqsecf=cmbx9 scaled 1000
  \font\secsy=cmsy9 scaled 1440
  \font\qsecsy=cmsy9 scaled 1200
  \font\qqsecsy=cmsy9 scaled 1000
  \textfont\secfam=\secf \scriptfont\secfam=\qsecf
  \scriptscriptfont\secfam=\qqsecf
  \def\secfont{\currfam\secfam \sqrfam\secfam \fam\secfam\secf
     \textfont2=\secsy \scriptfont2=\qsecsy \scriptscriptfont2=\qqsecsy}

\newfam\subfam
  \font\subf=cmsl9 scaled 1440
  \font\qsubf=cmsl9 scaled 1200
  \font\qqsubf=cmsl9 scaled 1000
  \textfont\subfam=\subf \scriptfont\subfam=\qsubf
  \scriptscriptfont\subfam=\qqsubf
  \def\subfont{\currfam\subfam \sqrfam\BIGfam \fam\subfam\subf
     \textfont2=\secsy \scriptfont2=\qsecsy \scriptscriptfont2=\qqsecsy}

\newfam\subsubfam
  \font\subsubf=cmti9 scaled 1440
  \font\qsubsubf=cmti9 scaled 1200
  \font\qqsubsubf=cmti9 scaled 1000
  \textfont\subsubfam=\subsubf \scriptfont\subsubfam=\qsubsubf
  \scriptscriptfont\subsubfam=\qqsubsubf
  \def\subsubfont{\currfam\subsubfam \sqrfam\BIGfam \fam\subsubfam\subsubf
     \textfont2=\secsy \scriptfont2=\qsecsy \scriptscriptfont2=\qqsecsy}

\def\sq{\fam\sqrfam}



\def\mathscript#1#2{\ifmmode\mathchoice{#1\scriptstyle{#2}}{#1\scriptstyle{#2}}
  {#1\scriptscriptstyle{#2}}{#1\scriptscriptstyle{#2}}
  \else $\fam=\currfam
  \mathchoice{#1\scriptstyle{#2}}{#1\scriptstyle{#2}}
  {#1\scriptscriptstyle{#2}}{#1\scriptscriptstyle{#2}}$\fi}
\def\savefam{\everymath={\fam\currfam}}

\def\puncspace{\ifmmode\,\else{\ifcat.\C{\if.\C\else\if,\C\else\if?\C\else%
\if:\C\else\if;\C\else\if-\C\else\if)\C\else\if/\C\else\if]\C\else\if'\C%
\else\if$\C\else\space\fi\fi\fi\fi\fi\fi\fi\fi\fi\fi\fi}%
\else\if\empty\C\else\if\space\C\else\space\fi\fi\fi}\fi}
\def\SP{\let\\=\empty\futurelet\C\puncspace}


\def\eg.{{\it e.\thinspace g.}}
\def\ie.{{\it i.\thinspace e.}}
\def\etal.{{\it et al.}}
\def\cf.{{\it cf.}}

\def\UGC{{\sq UGC\kern.2em}}
\def\NGC{{\sq NGC\kern.2em}}
\def\IC{{\sq IC\kern.2em}}
\def\M{{\sq M\kern.06em}}

\def\I{\mathscript{\mskip2.5mu}{I}\SP}
\def\II{\mathscript{\mskip2.5mu}{II}\SP}


\def\kms{{\sq\savefam\thinspace km\thinspace s\sup{-1}}\SP}

\def\degree#1{\ifmmode{\if.#1{{^\circ}\llap.}\else{^\circ} #1\fi}\else
{\if.#1$^\circ$\llap.\else\if\empty#1$^\circ$#1\else$^\circ$ #1\fi\fi}\fi}
\def\deg#1{\ifmmode{\if.#1{{^\circ}\llap.}\else{^\circ} #1\fi}\else
{\if.#1$^\circ$\llap.\else\if\empty#1$^\circ$#1\else$^\circ$ #1\fi\fi}\fi}
\def\arcmin#1{\ifmmode{\if.#1{'\llap.}\else{'} #1\fi}\else
{\if.#1$'$\llap.\else$'$ #1\fi}\fi}
\def\arcsec#1{\ifmmode{\if.#1{''\llap.}\else{''} #1\fi}\else
{\if.#1$''$\llap.\else$''$ #1\fi}\fi}
\def\sun{\ifmmode _\odot \else $_{\odot}$\fi\SP}
\def\earth{\ifmmode _\oplus \else $_{\oplus}$\fi\SP}

\def\ee#1{\ifmmode {} \times 10^{#1} \else ${} \times 10^{#1}$\fi}
\def\sub#1{\ifmmode _{\sq #1} \else $_{\sq #1}$\fi}
\def\sup#1{\ifmmode ^{\sq #1} \else $^{\sq #1}$\fi}

\def\about{\ifmmode \sim \else $\sim$\kern.03em \fi}
\def\aboutless{\ifmmode \simless \else $\simless$\kern.03em \fi}
\def\aboutmore{\ifmmode \simgreat \else $\simgreat$\kern.03em\fi}
\def\x{${}\times{}$}

 \mathcode`*="002A 
\def\simless{\mathbin{\lower 3pt\hbox
     {$\rlap{\raise 5pt\hbox{$\char'074$}}\mathchar"7218$}}} 
\def\simgreat{\mathbin{\lower 3pt\hbox
     {$\rlap{\raise 5pt\hbox{$\char'076$}}\mathchar"7218$}}} 



\def\center#1{{\def\\{\break}\rightskip=0pt plus1fil \leftskip=\rightskip
     \parindent=0pt \parfillskip=0pt #1\par}}

\def\newline{\hfil\break} 
\def\newpage{\vfill\eject}

\def\SMALLskip{\vglue\smallskipamount\nobreak}
\def\MEDskip{\vglue\medskipamount\nobreak}
\def\BIGskip{\vglue\bigskipamount\nobreak}

\catcode`\@=11
\def\opencases#1{\left\{\,\vcenter{\openup\jot\m@th
    \ialign{$##\hfil$&\quad##\hfil\crcr#1\crcr}}\right.}
\catcode`\@=12


\def\nopagenumber{\global\nopagenumbertrue}
\def\nopagenumbers{\global\nopagenumberstrue}
\def\pagenumbers{\global\nopagenumbersfalse \global\nopagenumberfalse}
\headline={\ifnopagenumbers{}
  \else\ifnopagenumber{}\pagenumbers
  \else\ifchapterpage{}
  \else\ifprelimpage{}
  \else{\hss\TXrm\folio\hss}\fi\fi\fi\fi}
\footline={\ifnopagenumbers{}
  \else\ifnopagenumber{}\pagenumbers
  \else\ifchapterpage{\hss\TXrm\folio\hss}\global\chapterpagefalse
  \else\ifprelimpage{\hss\TXrm\folio\hss}
  \else{}\fi\fi\fi\fi}


\def\makeheadline{\vbox to 0pt{\vskip-\vheadline
  \line{\vbox to\ht\strutbox{}\the\headline}\vss}\nointerlineskip}
\def\makefootline{\baselineskip=\vfootline \line{\the\footline}}

\catcode`\@=11
\def\vfootnote#1{\insert\footins\bgroup\footnotefont\baselineskip=\footbaseskip
  \interlinepenalty\interfootnotelinepenalty
  \splittopskip\ht\strutbox 
  \splitmaxdepth\dp\strutbox \floatingpenalty=20000
  \parfillskip=0pt plus1fil \parindent=25pt
  \leftskip=0pt \rightskip=0pt \spaceskip=0pt \xspaceskip=0pt
  \textindent{#1}\footstrut\futurelet\next\fo@t}
\catcode`\@=12
\skip\footins=24pt
\def\footnoterule{\kern-10pt \hrule width 2.5truein \kern 9.5pt}
\newcount\notenumber

\def\note{\global\advance\notenumber by 1 $^\the\notenumber$
     \vfootnote{$^{\the\notenumber}$}}

\newcount\eqnumber
\def\clreqnumber{\global\eqnumber=0} \clreqnumber
\def\EQN#1#2$${\global\advance\eqnumber by1%
  \eqno\hbox{\rm(\chaphead\the\eqnumber#1)}$$\def\name{#2}\ifx\name\empty%
  \else\xdef#2{(\chaphead\the\eqnumber#1)\noexpand\SP}\fi\ignorespaces}
\def\eqn#1$${\EQN{}{#1}$$}
\def\eqna#1$${\EQN{a}{#1}$$}
\def\eqnb#1$${\global\advance\eqnumber by-1 \EQN{b}{#1}$$}
\def\eqnc#1$${\global\advance\eqnumber by-1 \EQN{c}{#1}$$}
\def\lasteqn{{\rm(\chaphead\the\eqnumber)}\SP}
\def\nexteqn{\advance\eqnumber by1 {\rm(\chaphead\the\eqnumber)}\advance
  \eqnumber by-1 \SP}

\newcount\fignumber
\def\clrfignumber{\global\fignumber=0} \clrfignumber
\newdimen\figwidth \newdimen\figheight \figwidth=5truein \figheight=5truein
\def\figbox{\vbox to \figheight{{\parindent=0pt\offinterlineskip
  \hbox to\hsize{\hfil \hbox to\figwidth{\vrule width1pc height.2pt depth0pt
  \hfil \vrule width1pc height.2pt depth0pt} \hfil}
  \hbox to\hsize{\hfil \hbox to\figwidth{\vrule height1pc width.2pt depth0pt
  \hfil \vrule height1pc width.2pt depth0pt} \hfil} \vfil
  \hbox to\hsize{\hfil \hbox to\figwidth{\vrule height1pc width.2pt depth0pt
  \hfil \vrule height1pc width.2pt depth0pt} \hfil}
  \hbox to\hsize{\hfil \hbox to\figwidth{\vrule width1pc height.2pt depth0pt
  \hfil \vrule width1pc height.2pt depth0pt} \hfil}}}}
\def\figdim#1x#2 {\figwidth=#1truein \figheight=#2truein}
\def\fig#1[#2]{\global\advance\fignumber by1 \def\name{#1}\ifx\name\empty%
  \else\xdef#1{\chaphead\the\fignumber\noexpand\SP}\fi \pageinsert
  \hrule height0pt \vfill\figbox\MEDskip\narrower\baselineskip=\footbaseskip
  {\tenpoint\noindent{\bf Figure \chaphead\the\fignumber.}\quad #2}
  \par\vfill\endinsert}
\def\figcont{\advance\fignumber by-1\fig[{\it (Continued.)}]}
\def\lastfig{{\rm\chaphead\the\fignumber}\SP}
\def\nextfig{\advance\fignumber by1 {\rm\chaphead\the\fignumber}\advance
  \fignumber by-1\SP}

\def\titlepage{\hrule height 0pt \nobreak\vskip 0pt plus0.8fil
     \global\nopagenumber \normalbaselines}
\def\endtitlepage{\par \ifdim\lastskip<0pt plus 0.5fil\vskip0pt plus 1.3fil\fi
   \normalbaselines \eject}
\def\title#1{\vskip 0pt plus .1\vsize
     \center{\baselineskip 20pt \BIGskip \titlefont\savefam\ #1}}
\def\author#1{\center{\baselineskip 20pt \MEDskip \sc\ #1}}

\def\date#1{\center{\baselineskip 20pt \BIGskip \it\ #1}}
\def\and{{\baselineskip 20pt \MEDskip \center{\sc\ and}}}


\def\abstract{\BIGskip \begingroup \centerline{ABSTRACT} \leftskip=20pt
     \rightskip=20pt \baselineskip=14pt \parindent=15pt \nobreak\SMALLskip
      \parskip=0pt}


\newcount\secno \newcount\subsecno \newcount\subsubsecno
\secno=0 \subsecno=0 \subsubsecno=0
\def\chapno{} \def\chaphead{}
\def\chapter#1 #2\par{\CHAPTER{#1}{#2}{Chapter}}
\def\appendix#1 #2\par{\CHAPTER{#1}{#2}{Appendix}}
\def\prelim#1\par{\global\prelimpagetrue\CHAPTER{}{#1}{}}
\def\CHAPTER#1#2#3{\vfill\eject\global\chapterpagetrue
  \secno=0\subsecno=0\subsubsecno=0
  \ifnum\pageno=1\message{#3 #1 currently begins on page 1.
  Enter new first page number:}\read-1 to\pagebeg
  \pageno=\number\pagebeg\fi
  \BIGskip\BIGskip
  \def\chapno{#1.}\def\chaphead{#1--}\message{\noexpand{#3 #1 #2}}
  \center{{\baselineskip30pt\titlefont\savefam #3\ #1\\#2}}
  \BIGskip\noindent\ignorespaces}
\def\preface#1\par{\ifnum\pageno=1\pageno=-1\fi
  \vfill\eject\global\prelimpagetrue \secno=0\subsecno=0\subsubsecno=0
  \BIGskip\BIGskip \message{Preface}
  \center{{\baselineskip30pt\titlefont\savefam Preface\\#1}}
  \BIGskip\noindent\ignorespaces}
\def\section#1\par{\advance\secno by1 \subsecno=0 \subsubsecno=0
  \vskip0pt plus.05\vsize
  \medbreak\MEDskip\message{Section \the\secno}
  \center{{\baselineskip24pt\secfont\savefam\chapno\the\secno\quad#1}}
  \nobreak\MEDskip\noindent\ignorespaces}
\def\subsection#1\par{\advance\subsecno by1 \subsubsecno=0
  \vskip0pt plus.03\vsize\penalty-100 \medbreak
  \message{Subsection \the\subsecno}\center{{\baselineskip 24pt
  \secfont\chapno\the\secno.\the\subsecno\quad\subfont\savefam#1}}
  \nobreak\SMALLskip\noindent\ignorespaces}
\def\subsubsection#1\par{\advance\subsubsecno by1\vskip0pt plus.02\vsize
  \penalty-40 \smallbreak\message{Subsubsection #1}\center{{\baselineskip 24pt
  \secfont\chapno\the\secno.\the\subsecno.\the\subsubsecno\quad
  \subsubfont\savefam#1}}\nobreak\SMALLskip\noindent\ignorespaces}



\def\references{\vskip0pt plus.05\vsize \medbreak\MEDskip\message{References}
  \center {\secfont References}\nobreak\MEDskip\begingroup \frenchspacing
  \parskip10pt \baselineskip\footbaseskip \let\everypar=\filbreak
  \def\par{\endgraf\hangindent24pt}\parindent0pt \vskip-10pt}
\def\endreferences{\endgroup}

\def\figurecaptions{\newpage\nopagenumbers
  \vskip0pt plus.05vsize \medbreak\MEDskip\message{Figure Captions}
  \center {\secfont Figure Captions}\nobreak\MEDskip\begingroup
  \parskip8pt \baselineskip14pt \def\par{\endgraf\hangindent40pt}
\parindent0pt}

\def\address #1:#2\par{\vskip 0pt plus.2\vsize \BIGskip\BIGskip
     {\hangindent30pt\baselineskip14pt
      \noindent{\sc\ #1}:{\rm\ #2}\par}\ignorespaces}

\newcount\tablenumber \newbox\tabbox \newdimen\tabwid
\def\clrtablenumber{\global\tablenumber=0} \clrtablenumber
\def\lasttable{{\rm\chaphead\the\tablenumber}\SP}
\def\nexttable{\advance\tablenumber by1 {\rm\chaphead\the\tablenumber}%
   \advance\tablenumber by-1 \SP}
\def\PM{\ifmmode\pm\else${}\pm{}$\fi}

\def\trule{\noalign{\vskip4pt}\noalign{\hrule}\noalign{\vskip4pt}}

\let\tablefont=\tenpoint
\font\sc=cmcsc10
\newdimen\tindent \newskip\tbaseskip \tbaseskip=12pt

{\catcode`\?=\active \catcode`\"=\active \catcode`\^^M=\active%
\gdef\table#1[#2]{\global\advance\tablenumber by1\begingroup%
  \ifx#1\empty\else\xdef#1{\chaphead\the\tablenumber\noexpand\SP}\fi%
  \tablefont \baselineskip\tbaseskip%
  \def\hhh{\hskip0pt plus1000pt}\def\?{\char`\?}%
  \def\a{\rlap{$^a$}} \def\b{\rlap{$^b$}} \def\c{\rlap{$^c$}}%
  \def\d{\rlap{$^d$}} \def\e{\rlap{$^e$}} \def\f{\rlap{$^f$}}%
  \def\g{\rlap{$^g$}} \def\h{\rlap{$^h$}} \def\i{\rlap{$^i$}}%
  \def\j{\rlap{$^j$}} \def\k{\rlap{$^k$}} \def\l{\rlap{$^l$}}%
  \def\|##1{\rlap{$^{\rm ##1}$}}%
  \def\ms##1{\multispan ##1}%
\def\...{\null\nobreak\leaders\hbox to0.5em{\hss.\hss}\hskip1.5em plus1filll\
}%
  \catcode`\"=\active%
  \catcode`\^^I=4 \catcode`\^^M=\active \catcode`~=\active \catcode`?=\active%
  \def?{\enspace} \def~{\hfil} \def^^M{\crcr} \def"{\quad}\tabskip=0pt%
  \midinsert\vbox to\vsize\bgroup\hrule height0pt\vfil%
  \centerline{TABLE \chaphead\the\tablenumber}%
  \def\tablename{#2}\if\tablename\empty\else\center{{\sc\tablename}}\fi%
  \setbox\tabbox=\vbox\bgroup\halign\bgroup}%
\gdef\tabcont{\begingroup%
  \tablefont \baselineskip\tbaseskip%
  \def\hhh{\hskip0pt plus1000pt}\def\?{\char`\?}%
  \def\a{\rlap{$^a$}} \def\b{\rlap{$^b$}} \def\c{\rlap{$^c$}}%
  \def\d{\rlap{$^d$}} \def\e{\rlap{$^e$}} \def\f{\rlap{$^f$}}%
  \def\g{\rlap{$^g$}} \def\h{\rlap{$^h$}} \def\i{\rlap{$^i$}}%
  \def\j{\rlap{$^j$}} \def\k{\rlap{$^k$}} \def\l{\rlap{$^l$}}%
  \def\|##1{\rlap{$^{\rm ##1}$}}%
  \def\ms##1{\multispan ##1}%
\def\...{\null\nobreak\leaders\hbox to0.5em{\hss.\hss}\hskip1.5em plus1filll\
}%
  \catcode`\"=\active%
  \catcode`\^^I=4 \catcode`\^^M=\active \catcode`~=\active \catcode`?=\active%
  \def?{\enspace} \def~{\hfil} \def^^M{\crcr} \def"{\quad}\tabskip=0pt%
  \midinsert\vbox to\vsize\bgroup\hrule height0pt\vfil%
  \centerline{TABLE \chaphead\the\tablenumber{ \it (continued)}}%
  \setbox\tabbox=\vbox\bgroup\halign\bgroup}%
\gdef\tnotes{\trule\egroup\egroup\tabwid=\wd\tabbox%
  \hbox to\hsize{\hfill{\box\tabbox}\hfill}%
  \def\a{{$^a$}} \def\b{{$^b$}} \def\c{{$^c$}}%
  \def\d{{$^d$}} \def\e{{$^e$}} \def\f{{$^f$}}%
  \def\g{{$^g$}} \def\h{{$^h$}} \def\i{{$^i$}}%
  \def\j{{$^j$}} \def\k{{$^k$}} \def\l{{$^l$}}%
  \def\|##1{{$^{\rm ##1}$}}%
  \baselineskip\footbaseskip%
  \advance\tabwid by-\hsize\divide\tabwid by-2\advance\tabwid by20pt%
  \leftskip=\tabwid%
  \rightskip=\tabwid \parskip=1pt \parindent=0pt \catcode`\^^M=5%
  \def\par{\endgraf\leftskip=\tabwid\rightskip=\tabwid\hangindent=10pt}%
  \def\endtable{\par\vfil\egroup\endinsert\endgroup}\hangindent=10pt}}%
\def\endtable{\trule\egroup\egroup\hbox to\hsize{\hfill{\box\tabbox}\hfill}
  \vfil\egroup\endinsert\endgroup}

\topskip=16pt
\splittopskip=16pt
\twelvebaseskip=24.6		
\tenbaseskip=18pt		
\footbaseskip=16pt		
\let\footnotefont=\twelvepoint  
\twelvepoint                    
\smallskipamount=8pt plus1pt
\medskipamount=16pt plus2pt
\bigskipamount=32pt plus4pt
\parskip=0pt
\parindent=40pt
\hsize6.1truein
\vsize8.75truein
\hoffset0.0truein
\voffset0.0truein
\vheadline.45truein
\vfootline.55truein

\jot=16pt
\abovedisplayskip=10pt plus 3pt minus 6pt
\abovedisplayshortskip=0pt plus 3pt
\belowdisplayskip=10pt plus 3pt minus 6pt
\belowdisplayshortskip=6pt plus 3pt minus 6pt

\def\SP{\let\\=\empty\futurelet\C\puncspace}
\newif\ifFP
\def\footpunc{\FPtrue\ifcat.\C{\if.\C\else\if,\C\else\if?\C\else%
\if:\C\else\if;\C\else\if-\C\else\if--\C\else\if---\C%
\if)\C\else\if]\C\else\if'\C\else\if\empty\C\else\if\space\C\else\FPfalse%
\fi\fi\fi\fi\fi\fi\fi\fi\fi\fi\fi\fi\fi}\else\FPfalse\fi}

\def\etal.{et al.}

\newwrite\figcap
\immediate\openout\figcap=figcaptions
\def\figcont{\null}
\def\fig#1[{\global\advance\fignumber by1 \def\name{#1}\ifx\name\empty%
  \else\xdef#1{\\the\fignumber\noexpand\SP}\fi
  \immediate\write\figcap{}
  \immediate\write\figcap{\medbreak\hangindent\parindent
  \noindent{\bf Fig. \the\fignumber.}---\ignorespaces}\copytoblankline}
\gdef\copytoblankline{\begingroup\setupcopy\dospecials
  \catcode`\|=12 \obeylines \copycap}
\def\setupcopy{\def\do##1{\catcode`##1=12}}
{\obeylines \gdef\copycap#1
  {\def\next{#1}\ifx\next\empty\let\next=\endgroup%
  \else\immediate\write\figcap{\next} \let\next=\copycap\fi\next}}
{\catcode`\]=\active
\gdef\figurecaptions{\immediate\closeout\figcap\normalbaselines\begingroup
  \twelvepoint\catcode`\]=\active\let]=\null
  \newpage\nopagenumbers\BIGskip\centerline{{\secfont Figure Captions}}
  \message{Figure Captions}\medskip\input figcaptions \par\vfill\endgroup
  \break}}

\def\titlepage{\hrule height 0pt \nobreak\vskip 0pt plus0.2fil
     \global\nopagenumber \normalbaselines}
\def\endtitlepage{\par \ifdim\lastskip<0pt plus 0.5fil\vskip0pt plus 1.3fil
  \else\vskip0pt plus 0.8fil\fi
  \normalbaselines \eject
\gdef\note##1{\global\advance\notenumber by1{$^{\the\notenumber}$}\begingroup
  \abovedisplayskip=0.3\baselineskip\belowdisplayskip=0.1\baselineskip
  $$\overline{\underline
  {\hbox to\hsize{\vbox{\noindent$^{\the\notenumber}$##1}}}}$$\endgroup
  \ignorespaces}}

\def\title#1{\center{\baselineskip 20pt \BIGskip \titlefont\savefam\ #1}}
\def\author#1{\center{\baselineskip 20pt \MEDskip \sc\ #1}}

\def\date#1{\center{\baselineskip 20pt \BIGskip \it\ #1}}
\def\and{{\baselineskip 20pt \MEDskip \center{\sc\ and}}}


\def\abstract{\SMALLskip \vskip0pt plus0.6fil \begingroup\centerline{ABSTRACT}
  \leftskip=20pt \rightskip=20pt plus2em \parindent=15pt \nobreak\SMALLskip
  \parskip=0pt}


\def\section#1\par{\advance\secno by1 \subsecno=0 \subsubsecno=0
  \vskip0pt plus.05\vsize
  \medbreak\SMALLskip\message{Section \the\secno}
  \center{\baselineskip24pt\secfont\savefam
  \the\secno.\quad#1}
  \nobreak\MEDskip\noindent\ignorespaces}
\def\subsection#1\par{\advance\subsecno by1 \subsubsecno=0
  \vskip0pt plus.03\vsize\penalty-100 \medbreak
  \message{Subsection \the\subsecno}\center{\baselineskip 24pt
  \secfont\the\secno.\the\subsecno\quad\savefam\subfont#1}
  \nobreak\SMALLskip\noindent\ignorespaces}
\def\subsubsection#1\par{\advance\subsubsecno by1\vskip0pt plus.02\vsize
  \penalty-40 \smallbreak\message{Subsubsection #1}\center{\baselineskip 24pt
  \secfont\the\secno.\the\subsecno.\the\subsubsecno\quad
  \savefam\subsubfont#1}\nobreak\SMALLskip\noindent\ignorespaces}
\def\appendix#1 #2\par{\vfill\eject\secno=0\subsecno=0\subsubsecno=0
  \medbreak\MEDskip
  \center{{\baselineskip24pt\secfont\savefam Appendix #1.\quad#2}}
  \nobreak\MEDskip\noindent\ignorespaces}

\input /home/qedesh/edvige/tex/tablex.tex


\def\address #1:#2\par{\vskip 0pt plus.2fill \BIGskip
  {\hangindent30pt\noindent{\sc #1}:{\rm\ #2}\par}\ignorespaces}

\def\references{\vfill\eject\message{REFERENCES}
  \center {\secfont REFERENCES}\nobreak\MEDskip\begingroup \frenchspacing
  \parskip10pt \baselineskip\footbaseskip \let\everypar=\filbreak
  \def\par{\endgraf\hangindent24pt}\parindent0pt \vskip-10pt}
\def\endreferences{\endgroup}

\input /home/qedesh/edvige/tex/apjref.tex
\twelvebaseskip=18pt            
\tenbaseskip=18pt               
\footbaseskip=18pt              
\let\footnotefont=\twelvepoint  
\twelvepoint                    
\smallskipamount=6pt plus1pt
\medskipamount=12pt plus2pt
\bigskipamount=24pt plus4pt
\abovedisplayskip=6pt plus 3pt
\belowdisplayskip=3pt plus 3pt
\abovedisplayshortskip=3pt
\belowdisplayshortskip=0pt
\parindent=36pt
\parskip=0pt
\hsize6.5truein
\vsize9.0truein
\hoffset0.0truein
\voffset-0.0truein
\nopagenumber
\tolerance=2000
\hfuzz=10pt
\overfullrule=0pt
\titlepage

\def\uv{$u$--$v$}
\def\affl#1{\noindent\llap{$^{#1}$}}
\newdimen\windowhsize \windowhsize=13.1truecm
\newdimen\windowvsize \windowvsize=6.6truecm
\def\eol{\hfil\break}
\def\heading#1{
    \vskip0pt plus6\baselineskip\penalty-250\vskip0pt plus-6\baselineskip
    \vskip2\baselineskip\vskip 0pt plus 3pt minus 3pt
    \centerline{\bf#1}
    \global\count11=0\nobreak\vskip\baselineskip}
\count10=0
\nopagenumbers\parindent=0pt
\footline={\ifnum\pageno<1 \hss\thinspace\hss
    \else\hss\folio\hss \fi}
\pageno=-1

{
\def\cl#1{\hbox to \windowhsize{\hfill#1\hfill}}
\hbox to\hsize{\hfill\hbox{\vbox to\windowvsize{\vfill
\bf
\cl{NEUTRAL HYDROGEN ABSORPTION AND EMISSION}
\cl{IN THE QUASAR/GALAXY PAIR 3C275.1/\NGC4651}
\bigskip
\cl{Stephen E. Schneider$^{1}$ and Edvige~Corbelli$^{2}$}
\bigskip\rm
\cl{Preprint n.~11/93}

\vfill}}\hfill}}

\vskip5truecm
{\leftskip1.7truecm
\affl{1}Five College Astronomy Department, and
\eol
Dept.\ of Physics and Astronomy, Univ.\ of Massachusetts,
\eol
632 Lederle Tower, Amherst MA 01003 (USA)
\bigskip
\affl{2}Osservatorio Astrofisico di Arcetri,
\eol
Largo E.~Fermi 5, I-50125 Firenze (Italy)

\vfill
The Astrophysical Journal, in press, Sept. 1, 1993.
\vglue3truecm
}
\eject

\vglue\windowvsize plus 0pt minus \vsize
\heading{ABSTRACT}

3C275.1 and \NGC4651 make a particularly interesting quasar/galaxy pairing
because of the alignment of such a strong radio emitter behind
the outer H\I disk of a relatively undisturbed spiral galaxy.
This provides an
opportunity to study the spin-temperature characteristics of atomic
hydrogen at low column densities, in an apparently star-free
environment.
We previously reported a tentative detection of absorption against
the quasar based on VLA C--array observations;
we have now made more sensitive maps of the H\I emission from \NGC4651
with the VLA D--array, and we have attempted to confirm the weak H\I absorption
against the quasar at higher spatial and spectral resolution in VLA B--array.
The possible absorption feature against this quasar appears to be
weaker than we previously suspected, even though it seems
fairly clear that H\I emission is present close to the line of sight
to the quasar.
The weakness of the possible absorption seems also to confirm, conversely,
the trend found in previous observations that where strong absorption
lines are seen, the galaxies show evidence of disturbance.
The possible detection of (or limits on) absorption suggest that the
neutral gas in the outer disk is quite warm. We use the
absorption and emission measurements to set lower limits on
the combination of heating inputs outside the star-forming regions
of a disk galaxy and/or the intensity of the cosmic background
radiation around 100 eV.

\vglue\windowvsize plus 1fill minus \vsize
\eject

\pageno=1
\section INTRODUCTION

Absorption systems at low redshift are particularly interesting
because of the possibility of examining a galaxy's gaseous component
in emission as well as absorption, unlike the much more numerous
cases of absorption identified at high redshift.
However, due to the varying identification techniques applied at
different redshifts, it is not entirely clear how to relate
the evolution and type of absorbing systems seen at different
epochs (Wolfe 1990, Blades 1988, Morton, York, and Jenkins 1986).
The absorption systems normally attributed to galaxies
at intermediate redshifts are usually associated with the redshifted
Mg\II line, whereas at low redshifts the optical Ca\II line is usually
used to identify an absorption system and this requires a higher
total column density of gas.
In almost all cases at low redshifts where Ca\II absorption lines
have been found the absorption line is broad and the foreground
galaxy shows evidence of merging or tidal disturbance
(Carilli and Van Gorkom 1992; Bowen et al.\ 1991).
This may be just a selection effect that results from the
fact that it is only in such systems that high H\I column density
gas is thrown out to large galactrocentric distances, and
furthermore the disturbance makes the lines Ca\II/Na\I more
easily detectable by increasing their equivalent width.

Absorption lines arising clearly from undisturbed outer disks or
halos of low-redshift galaxies have not been identified. Today's
normal disk galaxies might be the final product of high redshift
systems which have shrunk or have lost most of their outer regions
during evolution (Charlton and Salpeter 1989). In this case some
faint gas might have been left in the halo or in the disk
several tens of kiloparsecs beyond their optical radius and
might be detectable through the 21-cm line of neutral hydrogen.

The ability to detect 21 cm lines in absorption generally requires a
relatively high column density of gas, comparable to the inner disk
column densities required to detect Ca\II absorption.
However, if the spin temperature of H\I is as low as 100 K, then
21 cm absorption can be detected also for $N\sub{H\I}\sim 10^{19}$
cm$^{-2}$.
This is why in Paper~I (Corbelli \&\ Schneider 1990)
we approached the problem of the gas in outer regions by searching for
21-cm absorption against radio sources which lay along lines of sight
passing close to galaxies.
Cataloged radio sources of adequate brightness were sufficiently
numerous that we could examine a relatively large sample
of 59 radio background sources at angular
separations of up to seven optical radii from cataloged galaxies.
We searched for 21-cm absorption using the Arecibo radio telescope.
Our survey was primarily sensitive to the sort of narrow-line absorption
that might be expected from undisturbed gas because,
since no optical absorption redshifts were previously known, we
could not easily eliminate the possibility of broad features being
masked by sidelobe emission within the beam or by bandpass irregularities.
In the end, we found no clear-cut cases of absorption, except for
one suspected very weak feature which is the subject of this paper.

The lack of a significant presence of cold H\I well outside of the
optical dimensions of spiral galaxies leaves us with three possibilities:
$(i)$ the faint gas extends much further out but it is mostly warm;
$(ii)$ there is a sharp transition between H\I and H\II beyond a
critical column density;
or $(iii)$ the gas is sharply truncated after a certain radius.
Accurate constraints on the spin temperature and very sensitive
emission searches would help in deciding which of these three
alternatives generally applies.
If case $(i)$ or $(ii)$ are correct, the data would give
important information on the soft X-ray background or on other possible
heating mechanisms outside the star forming region of disk galaxies.
In two recent papers, Corbelli and Salpeter (1992$a$, 1992$b$) have
in fact shown how one could use spin temperature and sensitive emission
searches to constrain the cosmic background below 500 eV.

We have therefore investigated further the one ``suspicious'' feature we had
found in Paper~I using the VLA.\note{The Very Large Array is a facility of the
National Radio Astronomy Observatory operated by Associated
Universities, Inc., under contract with the National Science Foundation.}
The preliminary results in Paper~I were based on a relatively short
integration in the VLA C--array, which showed a possible 21 cm absorption
in the 3C275.1/\NGC4651 pair.
However, neither that observation nor earlier ones made at Westerbork by
Warmels (1988) were sufficiently sensitive to detect H\I emission as far
out as the position of 3C275.1.
H\I had been detected only for $N\sub{H\I} \ge 10^{20}$ cm$^{-2}$, while
in some previously studied cases 21-cm absorption had been found
at lower column densities (Carilli and van Gorkom 1992).
To further study the H\I emission in \NGC4651, we therefore observed the
galaxy in VLA D--array, which gives us a higher sensitivity to lower
column densities of H\I than our earlier C--array observations.

In contrast to the other low-redshift cases of 21 cm absorption, where Ca\II
absorption was also found, \NGC4651 shows a regular H\I disk and no clear
evidence of companions.
Thus the presence of such a strong radio source near \NGC4651 presents a
unique opportunity to study the properties of gas in the outer disk of a
relatively undisturbed disk by its 21 cm H\I absorption.
\NGC4651 has sometimes been regarded as peculiar or disturbed because of
a faint ``jet'' (Sandage, Veron, and Wyndham 1965) extending from the galaxy.
In Fig.\ 1, we show this morphology with a B--band image of \NGC4651
obtained at the NOAO Kitt Peak 0.9 m telescope.
Fig.\ 1 (a) shows that the central high surface brightness regions of \NGC4651
look like a fairly normal Sc galaxy, while the ``stretch'' in panel (b)
emphasizes the peculiar linear feature dubbed a jet.
The linear feature is clearly inconsistent with any traditional notion of a jet
since it neither points back at the nucleus of the galaxy nor is it projected
along the minor axis of the galaxy nor does it produce detectable radio
continuum emission.
The morphology of the feature is also difficult to match to a tidal stream
given its straightness right up to the immediate proximity of the galaxy,
where it makes a sharp angle with respect to the outer spiral arms.
Moreover there is no indication of a disturbance in the H\I emission
associated with the feature (Warmels 1988; this paper), therefore it seems
more likely to be a foreground or background object seen in projection,
or some sort of dynamical feature---perhaps similar to the Magellanic
Stream seen edge-on---which is far enough beyond the galaxy's disk
that it causes no obvious disturbance.
This is a very curious feature, but it does not seem to play
any role in the current investigation.

Because of the weakness of the possible absorption detected with C--array
we also re-observed the galaxy in B--array, which allowed us to use both
higher spatial and spectral resolutions than before.
The higher spatial resolution of the B--array gave us a much better match
to the actual angular size of the two principal emission components of
the quasar, each of which has a flux density greater than 1 Jy, and these
observations had the further advantage of spatially ``filtering out''
almost all of the H\I emission (because of the lack of short \uv\
spacings).
The results of the B--array observations suggest that the H\I absorption is
weaker than we previously suspected, if it is present at all, even though
the new D--array observations now give us better evidence that H\I emission
is present at the positions surrounding that of the quasar.
Despite the uncertainties
on the presence of a weak 21-cm absorption line, the combination of
new emission and absorption measurements we present in this paper
leads to some interesting conclusions on the conditions of the gas outside
the optical disk of \NGC4651, and it has implications for the heating
inputs in outer regions of quite normal and undisturbed galaxies.

In \S2 and \S3, we present specifics of the absorption and emission
observations and the data reduction performed at the VLA.
We discuss first the original C--array observations, and then in turn the
B--array observations made to examine the absorption, and the D--array
observations of the galaxy's emission.
In \S4 we attempt to interpret the observations, specifically attempting
to reconcile the differences between B-- and C--array.
Finally, in \S5 we summarize our results and discuss the implications of
these measurements for the spin temperature of neutral hydrogen in the
outer disks of galaxies.

\vfill
\eject

\section VLA OBSERVATIONS OF H\I ABSORPTION IN 3C275.1/\NGC4651

Three sets of 21 cm spectral line observations were made at the VLA, in
the B--, C--, and D--arrays giving approximately 6, 22, and 60
arcsec resolution respectively.
Various resolutions were necessary because of the ranges of angular sizes
involved: \NGC4651 has an angular diameter of about 5 arcmin, while the
quasar 3C275.1 consists of two primary components a few arcsec in
diameter.
The C--array observations were made 1989 August 4 as part of a survey of
three possible quasar/galaxy absorption systems identified in Paper I, and
follow-up B--array observations of the absorption were made 1990 August 4.
Observations of the H\I emission were made 1991 March 26 in D--array.
We also attempted absorption measurements in D--array
(using observing parameters identical to those for the B--array observations),
but confusion with H\I emission from the galaxy prevented any definitive
interpretation of the results; we do not discuss these observations further.

In this Section and in the next one we describe the reductions
and results of each set of observations, deferring a comprehensive
discussion of the results to \S 4.

\subsection VLA C--Array

The C--array observations were centered on the position of 3C275.1 and
at the velocity of 700 \kms which is close to where absorption was
suspected based on our earlier Arecibo spectrum.
The integration consisted of two 35 minute blocks of time, separated by
\about3 hours to improve \uv\ coverage.
Twenty-seven antennas and two polarizations were used.
On-line Hanning smoothing with 128 spectral channels over a 1.56 MHz
bandpass yielded a velocity resolution of \about 2.6 \kms.
Standard amplitude, phase, and bandpass calibration were performed using
primary VLA calibrator 3C286, which was fortuitously only about 18\deg
away. The synthesized beam had a full-width at half maximum of \about
22 arcsec.

Because of limitations on the total bandpass vs.\ number of channels
available, only velocities spanning the low-velocity half of H\I emission
from the galaxy were observed.
We produced a continuum map from the average of the first and last
several spectral channels for the purpose of removing continuum emission
from the individual channels, but since the last channels were not line-
free, the subtraction was not ideal.
A small residual at the position of the H\I with a velocity near
\about800\kms resulted, but the effect of this on the
continuum-subtracted maps was minor since the H\I emission was quite weak
and several arcminutes removed from the position of the quasar.
Essentially identical results were obtained after using only the channels
from the line-free end of the spectrum, or when we performed subtraction
of the continuum emission in the \uv\ plane.

The integrated flux density from 3C275.1 was measured to be 2.96 Jy
based on channels from the line-free end of the spectrum, which is
in good agreement with our measurement at Arecibo (Paper~I).
H\I in \NGC4651 could be seen in emission at velocities near 640 \kms in
the vicinity of the quasar.
We looked at the spectrum of each pixel at position of the quasar and we
noticed a dip near this velocity.
In Fig.\ 2 we plot part of the spectrum integrated over a 18\arcsec
\x 18 \arcsec region centered on the quasar.
(Note that this spectrum appears slightly different from the preliminary
one published in Paper I due to small changes in the reduction method
and a simpler pixel-weighting scheme.)
A maximum apparent optical depth of 0.0067$\pm$0.0012 is found at
$v_{hel} = 643 \kms$
based on a formal analysis of the mean and rms away from the dip.
A second broader but weaker feature is also marginally visible at
636 \kms, with a formal depth of 0.0033$\pm$0.0014.
Spectra were also made using different \uv weighting, different
levels of cleaning, and from different halves of the recorded data,
all giving consistent dips near 640 \kms. No significant
variations in the bandpass calibration at this velocity appeared
to be present.

\subsection  VLA B--array

A year later, we reobserved this system with spatial and frequency
resolutions better optimized to study the characteristics of the
suspected absorption.
The quasar contains two radio-bright lobes (1.1 and 1.4 Jy to the
north and south respectively) and a fainter (0.2 Jy) central source
according to a high-resolution map of Stocke, Burns, and Christiansen (1985).
The lobe sources are separated by 15\arcsec, and they are each several
arcsec across.
The central source is unresolved.

The B--array resolution of 6\arcsec was a good match to the size of the
two strong continuum sources, and it offered the possibility of
distinguishing between absorption characteristics at the two positions.
The observations were again centered on 3C275.1 at a heliocentric
velocity of 640 \kms.
The integration time was spread over a 10 hour observing run, with an
actual integration time of \about7 hours on source after accounting for
calibration and slewing.
Amplitude, phase, and bandpass calibration were again performed using
3C286.
Twenty-seven antennas and two polarizations were used.
On-line Hanning smoothing with 64 spectral channels over a 0.39 MHz
bandpass (\about80 \kms) yielded a velocity resolution of \about 1.3
\kms.
The synthesized beam had a full-width at half maximum of \about 6 arcsec.

No H\I emission was apparent in our channel maps, presumably because
of the combination of poorer surface-brightness sensitivity in the
B--array, and the loss of short \uv\ spacings,
but 3C275.1 was clearly resolved into its northern and southern (combined
with the central)
components with measured flux densities of approximately 0.9 and 1.9 Jy
respectively.
The integrated flux density measured from 3C275.1 was 2.94 Jy, in
excellent agreement with the C--array results.

When smoothed to the resolution of the C--array (both spatially, and
in frequency), the B--array results show much weaker evidence of
H\I absorption (Fig.\ 2).
The feature at 643 \kms has almost completely disappeared, leaving
only a weak possible absorption line 636 \kms.
Somewhat better evidence of absorption (Fig.\ 2) at 636 \kms was found in
the full resolution B--array data at the position corresponding to the
central source as mapped by Stocke et al.\ (1985).
The spectrum showed a slight curvature (of a few tenths of a percent)
so that a second-order polynomial fit was subtracted from the spectrum
shown in Fig.\ 2;
this did not significantly affect the estimates of optical depth.
The peak optical depth was $0.0059\pm0.0016$,
and the integrated optical depth was $\int\tau\,dv = 0.030\kms$.
Only a very slight suggestion of absorption at the 643 \kms velocity was
found when examining points aligned with the northern component of
3C275.1.

\subsection VLA D--array: H\I emission from \NGC4651

Additional observations of \NGC4651 were made to map out its H\I
emission properties.
D--array observations provide better H\I surface density sensitivity,
but there is also a more difficult problem in determining the emission
at a particular point because of the larger beam size.

The observations were centered on \NGC4651 at its systemic
heliocentric velocity of 800 \kms.
Approximately 6 hours of integration time for the emission measurement
was spread over a 10 hour observing run.
Amplitude, phase, and bandpass calibration were again performed using
3C286.
Twenty-seven antennas and two polarizations were used.
On-line Hanning smoothing with 64 spectral channels over a 3.125 MHz
bandpass (\about650 \kms) yielded a velocity resolution of \about 10
\kms.

The line-free channels to either side of the velocity range of the galaxy
were averaged and subtracted from the remaining channels, and the
resulting channel maps had a typical rms noise per beam of about 1 mJy.
The four times longer integration, and the lower velocity and spatial
resolutions of these maps made them about 20 times more sensitive to
extended emission than the original C--array observations, with a limiting
sensitivity (3$\sigma$) of about $10^{19}$ cm$^{-2}$.
The integrated flux density of 3C275.1 was measured at 2.70 Jy.

The integrated emission maps and flux-weighted mean velocity maps are
shown in Fig.\ 3.
Overall, the emission is fairly symmetric and the velocity field is quite
regular. Around the position of 3C275.1, the maps become somewhat irregular
because of small uncertainties in the bandpass correction which leave
significant residuals when such a strong continuum source is subtracted from
the individual channel maps.
Several different methods (including \uv line subtraction and \uv
source subtraction) were also attempted, but they all produced
equivalent results.
No indication of the optical ``jet'' is apparent in these maps.

The amount of H\I emission at the position of 3C275.1 was estimated from a
spectrum taken through the 20 pixels (the beam area was 18.6 pixels in our
image) centered at the position of the quasar in the
CLEANed image (Clark 1980).
Because of the difficulties in removing such a strong source,
a second-order baseline had to be fitted and subtracted in order to
estimate the line flux at the position of the quasar.
The resulting spectrum (Fig.\ 4) peaks at approximately 634 \kms in
agreement with the velocity of the possible absorption found in the
B--array observations.
To check that the baseline subtraction was reasonable, we also measured
the H\I spectrum in a narrow annulus surrounding the quasar.
Here the baseline was linear, and the resultant spectrum is in good
agreement.
The integrated flux over the line is 0.23 Jy \kms per beam in this
measurement, corresponding to an average column density of H\I over the
beam of 7.4\ee{19} cm$^{-2}$.

A column density measured with a beam this large represents only the average
amount of gas that might be in front of the quasar.
If the gas is clumpy, there may be more or less H\I along the particular
line of sight to the quasar. Fortunately, the quasar is fairly extended,
so it should more nearly sample the average properties of the gas.
However a systematic effect arises because of the exponential gradient
of H\I across the beam, which tends to bias the measurement toward the
higher values of the column density at the edge of the beam.
We estimate the correction to our measurement by modeling the
H\I as having an exponential decline with a scale length $s$, and the
beam as being a Gaussian with a half-power beam width
$\theta = \sqrt{8\ln2} a$ where $a$ is the standard deviation of the Gaussian.
The measured flux is based on the product of the beam and exponential decline:
$$\int e^{-{x^2\over2a^2}}\times e^{-{x\over s}} dr
= \int e^{-{(x-a^2/s)^2\over2a^2}} \times e^{a^2\over s^2} dr
= \sqrt{2\pi} a \times e^{a^2\over s^2},$$
which is a factor of $e^{a^2/s^2}$ greater than the
actual flux at that position.
Note that we can treat this as an essentially one-dimensional problem because
the beam is small compared to the curvature in iso-density contours
at the position of the measurement.
By measuring the fall off at corresponding positions in the other
three quadrants of the galaxy, we estimate the scale length of the
H\I at the position angle of the quasar as $s\approx 30''$.
For the beam width of $\theta = 60''$, the flux would be overestimated by
factor of \about2.1.
Thus the column density of H\I is better estimated as
\about3.5\ee{19} cm$^{-2}$ at the position of the quasar.

\section INTERPRETATION OF THE OBSERVATIONS

It is well-known that detecting weak spectral features in the presence
of strong continuum sources is difficult (see Van Gorkom and Ekers 1989,
Cornwell, Uson, and Haddad 1992, for a discussion of this problem),
but despite repeated attempts and different
methods of analysis, we can find no clear explanation for the
significantly stronger absorption we find with our C--array observations.
The most disturbing element of these observations is that
the earlier C--array feature was at least nominally
better than \about5-$\sigma$, while at the same
velocity the B--array shows no feature.
There does appear to be a lower-velocity absorption line which
might be consistent in both B-- and C--arrays,
but the feature is weak ($\tau_{max} \simless 0.006$), is limited
to regions near the center of 3C275.1, and is uncomfortably near
our detection limit.

The differences between B-- and C--array results could have been caused
by imperfect sidelobe subtraction of the H\I emission.
This can occur when negative sidelobes from H\I emission very near to
the quasar in a particular channel are imperfectly subtracted.
Yet after applying various taperings in the \uv plane, and
a variety of ``cleaning'' techniques, we cannot confirm this speculation.

For the variation to be real would require some improbable circumstances.
Motion transverse to the line of sight due either to rotation or to
turbulence in the disk of \NGC4651 is extremely unlikely to proceed rapidly
enough to cause the sorts of large changes we see.
Briggs (1988) has pointed out that a variable background
source may cause a shift in the position of the absorption line.
Since the overall fluxes measured with B-- and C--array are in very
good agreement,
we might suppose that the quasar has ejected a ``blob'' while maintaining
the same total flux, so that different lines of sight in \NGC4651
are effectively sampled.  However for the case we are examining this
explanation is effectively ruled out by the following argument:
$(i)$ most of the flux comes from the north and south lobes and not
from the central source which should be the source of such rapid motion;
and $(ii)$ 3C275.1 is at a redshift \about150 times greater than \NGC4651,
so that even assuming superluminal
motion for the blob in the quasar of $v\sim 10c$  the opacity in
the medium of \NGC4651 would have to change over scales
$\ll$1 pc to explain the observed variations in the absorption profiles.

For the time being, we attribute the apparently strong absorption
at 643 \kms to a random deviation in our C--array data.
This explanation is not entirely satisfactory, but until better data are
obtained, it seems a more reasonable alternative than invoking a change
in the geometry of the background source.
The weaker feature at 636 \kms appears to be
consistent in both the B-- and C--array observations and the peak
corresponds closely to the velocity of the emission maximum as determined
from the D--array observations.  However its weakness (it is a quarter
as strong as the feature in the C--array observation at 643 \kms)
precludes any definitive conclusion and in the following section we
will treat it as an upper limit.

\section DISCUSSION

Although some uncertainties remain about the strength of absorption
in the 3C275.1/ \NGC4651 pair, it now appears fairly clear that
H\I emission is present along the line of sight to the quasar from
the outer disk of \NGC4651 at the level of
$N\sub{H\I}\sim 3\times 10^{19}$ cm$^{-2}$.
The gas in this outer region appears to be relatively undisturbed,
making this alignment especially interesting for studies of the physical
characteristics of disk gas outside the stellar disk.
This should be contrasted with the majority of other large-separation
quasar/galaxy absorption systems where the 21-cm absorption appears to
be related to tidal features.
The weakness of the absorption in this system
is consistent with the idea that 21-cm absorption as well as
Ca\II or Na\I absorption are found prevalently in disturbed systems
which are richer in metals and have a higher neutral column density of
gas.
However even in this type of system it seems very likely
that if there is some cold gas, which absorbs the 21-cm radiation,
this is mixed with warmer gas.
It would be quite informative to explore Ca\II or Na\I absorption against
3C275.1, but unfortunately its optical counterpart is quite faint: $m_v =
19.0$ (Burbidge et al. 1971).
Extrapolating from observations like those of Bowen et al.\ (1991),
detection of the optical lines would require very lengthy integrations.
HST observations of the UV absorption spectrum might be a better
alternative if 3C275.1 is sufficiently bright at shorter wavelengths.

Independently of whether there is an absorption line in the
spectrum of 3C275.1 caused by gas associated with \NGC4651, we can
use the weakness of the absorption as a probe of possible heating inputs in
the outer regions of this low-redshift spiral galaxy.
Assuming the deeper feature observed with C--array is a ``glitch''
in the spectrum, and treating the apparent absorption at 636
\kms as either a detection or an upper limit to the optical depth ,
we have
$$T_S = 5.14\ee{-19} {N\sub{H\I}\over\int \tau\, dv} \approx 500 {\rm
K}$$
where we used the the column density from our D--array observations
and the integrated optical depth in front of the quasar from
our B--array data.
We note again that all absorption/emission measurements suffer from the
possibility that the gas could be clumpy. Thus the particular
column density along the line of sight to the quasar could be
higher or lower than the mean value represented by the emission in the
surrounding regions.

Applying a correction factor for the inclination of the galaxy with
respect to our line of sight, we find the H\I surface density in \NGC4651
at the location of the quasar: $\sigma\sub{H\I}\sim 2\times 10^{19}$
cm$^{-2}$.
Corbelli and Salpeter (1992$a$) have shown that such H\I
surface densities in outer regions
have spin temperatures $T_S>500$K only if the heat per atom
deposited by some mechanism is larger than $10^{-15}$ eV s$^{-1}$
(the exact value depends upon the gas volume density). If this
heating is provided by a cosmic flux described by the power law
$I\times E^{-2.4}$ ph cm$^{-2}$ s$^{-1}$ sr$^{-1}$ keV$^{-1}$ down
to 100 eV, then $I>10$. This result, based on half of solar metallicity,
does not depend much on the metallicity since in order to have these large
values of $T_S$ the medium should be in the warm phase with the cooling
dominated by H--e impact. The heating requirements do not change much even
if it is only one part of the gas which is cold and
absorbs the radiation. Given a total HI surface density of the
gas $\sim 2\times 10^{19}$ cm$^{-2}$, Corbelli and Salpeter
(1992$a$, Fig.\ 5) show in fact that a warm phase at $T_S>500$ K can
coexist with a cold one only if $I>10$. A two phase medium
would have been also the most logical conclusion derived from the stronger
feature previously thought to be present in the C--array data (due to its
small width and small optical depth).

These limits on the background flux around 100 eV are sufficiently
low that they have a direct bearing on the level of
a QSO-dominated background.
Madau (1992) estimated an intensity only two times lower than our
limit for a perfectly transparent intergalactic medium.
Due to possible rms
deviations and sidelobe contamination in the emission and absorption
spectra of 3C275.1/\NGC4651 we cannot exclude that such background
is the only heating source for the gas. A stronger background,
as recently suggested by new estimates of the bright QSO space density,
would give gas spin temperatures in even closer agreement with our estimates.
We would, however, exclude much weaker cosmic backgrounds unless
additional sources of energy are present in the outer regions of
normal spirals.
This conclusion is also supported by the consistency between the
values or lower limits for $T_S$ derived in this paper for the
outer disk of \NGC4651, and the lower limit of 250 K found for $T_S$
in the outer disk of M33 (Paper~I) for similar values of the H\I
surface density.

Although some doubts remain about weak features in outer regions, our
study is consistent with the conjecture that strong absorption
lines at low redshifts arise mainly in disturbed systems. At the
same time, because tidal features often have stars associated with
them and relatively high column densities of H\I, they are poorer
probes of the soft X-ray background. Even though the absorption features
may be weak and difficult to detect, we believe further studies of H\I
emission and absorption in outer reaches of undisturbed galaxies hold
great promise for exploring the energetic environments of galaxies.

We would like to thank E.\ Salpeter for a number of helpful discussions
during the preparation of this paper, J.\ Spitzak for helping to
obtain the B--band image of \NGC4651, and the referee for comments
on the original manuscript. This work was supported in part by a
PYI award from the NSF (AST-9158096) and the Agenzia Spaziale Italiana.
SES also thanks the Arecetri observatory for their hospitality and
support while visiting.

\references

Blades, J. C. 1988, in QSO Absorption Lines, eds. J. C. Blades et al.
(Cambridge University Press, Cambridge), p. 147

Bowen, D. V., Pettini, M., Penston, M. V., and Blades, J. C. 1991, MNRAS 249,
145.

Briggs, F. H. 1988, in QSO Absorption Lines, eds. J. C. Blades et al.
(Cambridge University Press, Cambridge), p. 275.

Burbidge, E. M., Burbidge, G. R., Solomon, P. M., and Strittmatter, P. A.
1971, ApJ, 170, 233

Carilli, C. L., and van Gorkom, J. H. 1992, ApJ, in press.

Charlton, J. C., and Salpeter, E. E. 1989, ApJ, 346, 101.

Clark, B. G. 1980, A\& A, 89, 377.

Corbelli, E. and Salpeter, E. E. 1992a, submitted to ApJ.

Corbelli, E. and Salpeter, E. E. 1992b, submitted to ApJ.

Corbelli, E., and Schneider, S. E. 1990, ApJ, 356, 14 (Paper~I).

Cornwell, T. J., Uson, J. M., and  Haddad, N. 1992, A\& A, 258, 583.

Madau, P. 1992, ApJ, 389, L1.

Morton, D. C., York, D. G., and Jenkins, E. B. 1986, ApJ, 302, 272.

Sandage, A. R., Veron, P., and Wyndham, J. 1965, ApJ, 142, 1307

Stocke, J. T., Burns, J. O., and Christiansen, W. A. 1985, ApJ, 299, 799.

Van Gorkom, J. H., and Ekers, R. D. 1989, in Synthesis Imaging in Radio
Astronomy, eds. R. A. Perley et al. (NRAO, Virginia), p. 341

Warmels, R. H. 1988, A\& AS, 72, 427

Wolfe, A. M. 1990, in The Interstellar Medium in Galaxies, eds. H. A.
Thronson and J. M. Shull (Kluwer, Dordrecht), p.387.

\endreferences

\vfill\eject

\center{Figure Captions}

Fig.\ 1.---{\it (a)} B--band image of \NGC4651 showing the relative position
of 3C275.1 (marked with arrow). {\it (b)} The same image ``stretched'' to
bring out the low surface brightness linear feature.

Fig.\ 2.---{\it (a)} Absorption spectrum against the entire quasar 3C275.1 as
found in the
C--array (dashed line) and B--array (solid line). The B--array data has been
smoothed both spatially and in frequency resolution to match the C--array.
The vertical scale is displayed relative to the total flux density over
the pixels sampled.
{\it (b)} Possible absorption feature found in B--array data as measured near
the center of 3C275.1.

Fig.\ 3.---{\it (a)} Contours of integrated H\I emission in \NGC4651 at column
densities corresponding to approximately
1.5, 3, 8, 15, 30, 80, 150, and 240 $\times 10^{19}$ cm$^{-2}$.
The data were taken in D--array with an effective beam size of
$62''\times59''$.
The beam size is shown as a cross-hatched region centered on the position
of the quasar.
{\it (b)} Isovelocity contours of the H\I, based on the flux-weighted mean,
at intervals of 20 km/s. The lowest contour, at left-hand side of the
figure, is for 630 km/s, the highest one is 950 km/s at the opposite
side of the figure.

Fig.\ 4.---H\I spectrum at the position of the quasar in D--array. The
solid line shows a baseline-subtracted spectrum in the direction of
3C275.1; the dashed line shows the spectrum from an annulus surrounding
the quasar.

\end